\documentclass[aps,prl,notitlepage,reprint,twocolumn,showpacs,superscriptaddress,longbibliography,nofootinbib,floatfix]{revtex4-1}

\usepackage{amsmath,amssymb,amsfonts}
\usepackage{graphicx}
\usepackage{verbatim}
\usepackage{pifont}

\usepackage{hyperref}
\usepackage{slashed}
\usepackage{verbatim}

\usepackage[usenames]{color}
\definecolor{darkgreen}{rgb}{0.1,0.65,0.1}

\newcommand{\be}{\begin{equation}}
\newcommand{\ee}{\end{equation}}
\newcommand{\bi}{\begin{itemize}}
\newcommand{\ei}{\end{itemize}}
\newcommand{\bea}{\begin{eqnarray}}
\newcommand{\eea}{\end{eqnarray}}
 
\newcommand{\bra}[1]{\langle\,#1\,|}          
\newcommand{\ket}[1]{|\,#1\,\rangle}          
\newcommand{\ud}{\mathrm{d}}

\newcommand{\LCm}{{\scriptscriptstyle -}}

\usepackage[T1]{fontenc} 

\begin{document}

\title{Quantum radiation reaction: from interference to incoherence}
\author{Victor Dinu}
\email{dinu@barutu.fizica.unibuc.ro}
\affiliation{Department of Physics, University of Bucharest, P.O.~Box MG-11, M\u agurele 077125, Romania}
\author{Chris Harvey}
\email{christopher.harvey@chalmers.se}
\affiliation{Department of Physics, Chalmers University of Technology, SE-41296 Gothenburg, Sweden}

\author{Anton Ilderton}
\email{anton.ilderton@chalmers.se}
\affiliation{Department of Physics, Chalmers University of Technology, SE-41296 Gothenburg, Sweden}
\author{Mattias Marklund}
\email{mattias.marklund@chalmers.se}
\affiliation{Department of Physics, Chalmers University of Technology, SE-41296 Gothenburg, Sweden}

\author{Greger Torgrimsson}
\email{greger.torgrimsson@chalmers.se}
\affiliation{Department of Physics, Chalmers University of Technology, SE-41296 Gothenburg, Sweden}

\begin{abstract}
We investigate quantum radiation reaction in laser-electron interactions across different energy and intensity regimes. Using a fully quantum approach which also accounts exactly for the effect of the strong laser pulse on the electron motion, we identify in particular a regime in which radiation reaction is dominated by quantum interference. We find signatures of quantum radiation reaction in the electron spectra which have no classical analogue and which cannot be captured by the incoherent approximations typically used in the high-intensity regime. These signatures are measurable with presently available laser and accelerator technology.
\end{abstract}
\maketitle
Intense light sources offer new prospects for observing quantum effects in laser-matter interactions. Phenomena such as particle beam spreading~\cite{Green:2013sla}, cooling~\cite{Neitz,Yoffe} and trapping~\cite{Gonoskov:2013aoa,Pukhov} can all be phrased in terms of the quantum recoil experienced by particles interacting with laser pulses, recoil which dominates particle motion in certain regimes~\cite{RDR}. Because of this the topic of quantum recoil, also called quantum radiation reaction (``QRR''), now receives a great deal of attention~\cite{DiPiazza:2011tq,Li,Vranic,Blackburn:2015tva,Zepf,Lisbon}.

Investigations of QRR often focus on high-intensity regimes currently out of experimental reach. In such regimes QRR comes from multiphoton emission, and the shortness of the `formation length' of quantum processes at high intensity implies that these emissions can be described as incoherent events~\cite{RitusReview,DiPiazza:2010mv}. In this Letter we show that the nature of QRR varies significantly in different intensity and energy regimes,  in particular regimes which are relevant to experiments soon to be performed.  In particular we reveal a regime, accessible with the laser intensities and accelerator technology available today, in which QRR is dominated by {\it coherent} quantum effects with no classical analogue, effects which are distinct from those in the high-intensity regime and which cannot be described by the approximations or numerical methods used there. Further, we will find new kinematic delineations of the different regimes.

Consider an electron interacting with a strong electromagnetic field. The classical Lorentz force equation predicts that the electron moves with some momentum~$\pi_\mu$. A measurement of the electron momentum would however yield a different result $P_\mu$, because the Lorentz equation does not account for the fact that the electron radiates and, by conservation of momentum, recoils when it does so~\cite{LAD}. The impact of this {\it radiation reaction} (``RR'') on the motion of the electron can be characterised simply by the difference between the actual momentum of the electron and that predicted by the Lorentz force: $P_\mu - \pi_\mu$ is classical RR. The momentum $P_\mu$ can be obtained as the classical or low-energy limit of a quantum mechanical observable, namely the expectation value of the electron momentum operator $\hat{P}_\mu$~\cite{Krivitsky:1991vt,Higuchi:2002qc,Higuchi:2004pr,Ilderton:2013tb}. Hence $\langle \hat{P}_\mu \rangle - \pi_\mu$ is a measure of QRR. The expectation value $\langle \hat{P}_\mu \rangle$ can be calculated for arbitrary weak fields in perturbation theory~\cite{Krivitsky:1991vt} but this is not sufficient for our purposes as the fields of interest are strong. In order to account fully for the impact of a strong laser field on electron motion, as well as giving a fully quantum treatment of $\langle \hat{P}_\mu\rangle$ in QED, we begin with a plane wave laser model. This is satisfactory in the high-energy regime we consider first, while beam focussing at high-intensity will be accounted for below. The QED calculation of $\langle \hat{P}_\mu\rangle$ follows~\cite{Ilderton:2013tb} and is described in the Supplementary~\cite{Supp}.

QRR effects depend on the following parameters. Let $\omega$ and $k_\mu$ be typical laser frequency and momentum scales, and let $p_\mu$ be the initial electron momentum.  Then the energy scale of the interaction is $b_0 \equiv k\cdot p/m^2$ which is $\simeq 2\omega\gamma/m$ for large~$\gamma$. (We use units such that $\hbar=c=1$ throughout.) Quantum effects in a field $F_{\mu\nu}$ are often characterised using the ``quantum efficiency parameter''~$\chi = \sqrt{p\cdot(eF)^2\cdot p} / m^3$. For a wave of  intensity $a_0 = eE/m\omega$, field strength~$E$, $\chi$ becomes the product $\chi = a_0 b_0$~\cite{RitusReview}. Hence a given $\chi$ may be achieved through different intensity/energy combinations, and we will see that different choices lead to very different physics.  We take the laser to propagate in the $z$-direction and be polarised in the $x$-direction, so that the laser fields depend on the phase $\phi \equiv \omega (t+z)$ through a potential with $x$-component $e A^x =  m a_0 e^{-\phi^2/\tau^2} \sin(\phi)$. We fix the wavelength at $\lambda = 2\pi/\omega = 820\,\text{nm}$ and choose $\tau$ such that the FWHM pulse duration is $15\,\text{fs}$.
  
\begin{figure}[t!]
	\includegraphics[width=0.95\columnwidth]{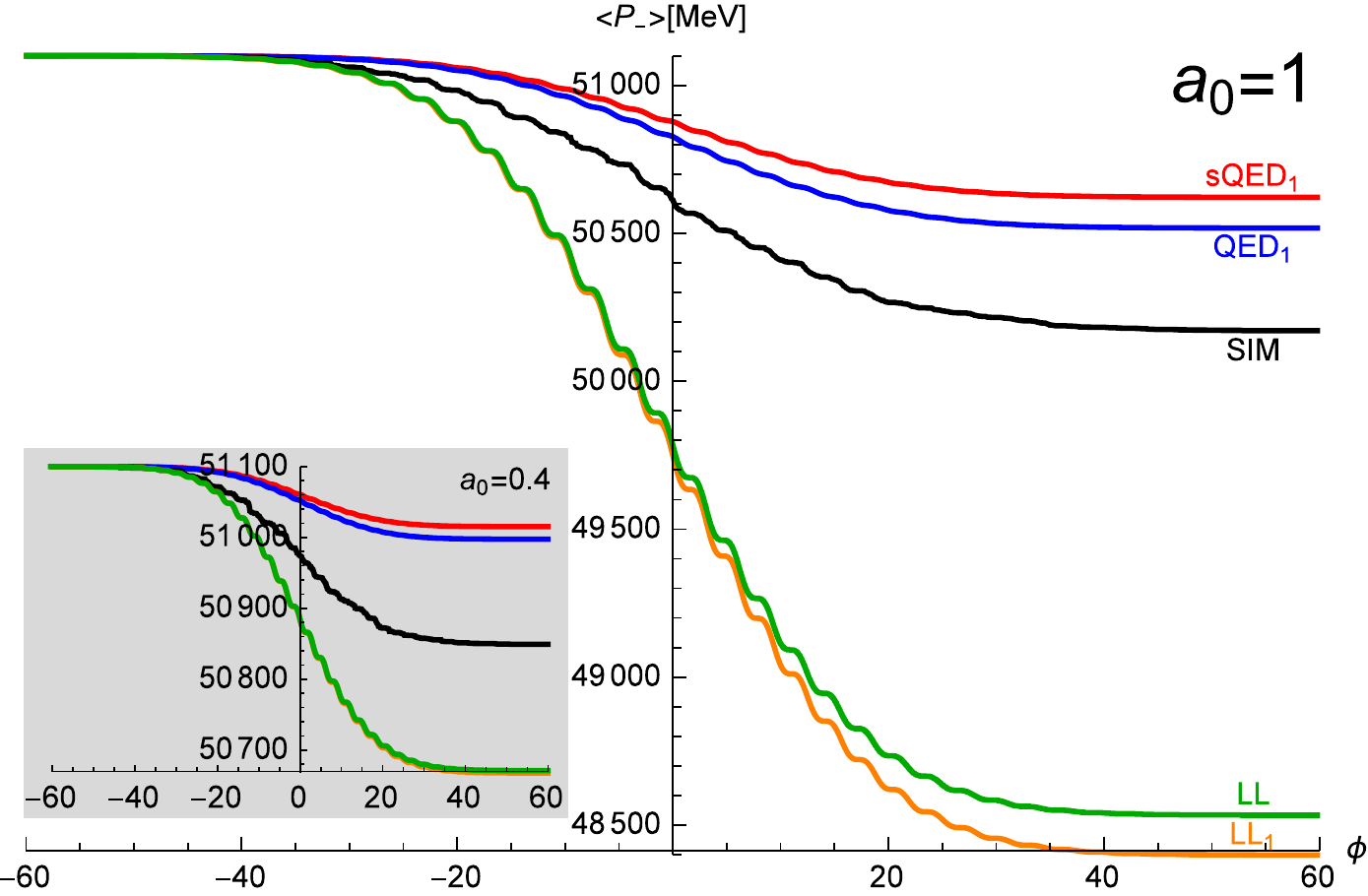}
	\caption{\label{FIG:SLAC} QRR in the interference-dominated regime, $a_0=1$ and $\gamma=10^5$. {\it Red/blue:} scalar QED/QED exact to order~$\alpha$ (subscript $1$). {\it Black/``SIM'':} simulation data based on the LCA. {\it Green:} exact solution of the classical LL equation. {\it Orange}/``LL$_1$'': the classical limit of the first order QED results. Inset: curves for $\gamma=10^5$, $a_0=0.4$, $\chi=0.24$.} 
\end{figure}
We begin our investigation with achievable parameters $a_0 = 1$, corresponding to an intensity of~$\sim10^{18}$~W/cm$^2$, and $\gamma = 10^5$~\cite{Bamber:1999zt}, suggesting a maximum $\chi=b_0=0.59$. In Fig.~\ref{FIG:SLAC} we plot, for a head-on collision, the electron momentum component $P_\LCm \equiv(E - p_z)/2$, the difference between energy and $z$-momentum, which shows the most significant deviation from the Lorentz-force result; $P_\LCm$ is conserved without RR, but recoil effects break this symmetry~\cite{Exact,Harvey:2011dp}. It is convenient to consider $\langle \hat{P}_\mu\rangle$ as a function of phase~$\phi$, as this relates the momentum to the local intensity in the laser pulse. (Collision at $45^\circ$ incidence, as may be experimentally necessary, can be advantageous as it makes QRR visible in all momentum components; for examples see the Supplementary~\cite{Supp}.)

Note first that Fig.~\ref{FIG:SLAC} shows only a small (5\%) difference between the exact solution of the classical Landau-Lifshitz (``LL'') equation~\cite{Exact} and the classical limit of the QED result (giving the first order solution of the LL and LAD equations~\cite{Krivitsky:1991vt,Higuchi:2002qc,Higuchi:2004pr,Ilderton:2013tb}). This suggests that higher-order multiphoton effects are small. However, classical predictions are invalid here: accounting for quantum effects clearly shows that the classical theory greatly overestimates RR losses, the relative error being around $350\%$. The inset in Fig.~\ref{FIG:SLAC} shows that quantum effects persist even for smaller $\chi$~\cite{Mironov:2014xba}.  Our QED approach allows us to account fully for spin, and the figure shows that spin slightly increases radiative losses relative to those in scalar QED~\cite{Dumlu:2011rr}. Fig.~\ref{FIG:SLAC} also shows results from by-now standard numerical simulations of intense laser-matter interactions which assume Lorentz force propagation between quantum emissions described in a locally constant approximation (``LCA'')~\cite{Elkina:2010up,Ridgers,Green:2014kfa,Gonoskov:2014mda}. The approximations behind the codes hold only for $a_0\gg1$, so they should not be expected to recover QRR in the considered regime; indeed the simulation data in Fig.~\ref{FIG:SLAC}, obtained from~$10^4$ runs, 
 fails to fully capture quantum effects.

To understand these results, in particular the {\it quantum reduction} of energy loss due to RR, we examine the structure of the average momentum $\langle \hat{P}_\mu \rangle$. To first order in $\alpha$ (the fine structure constant) and exactly in all other parameters, $\langle \hat{P}_\mu \rangle$ may be written, for an arbitrary pulse shape and duration, as
\be\label{P}
	\langle {\hat P_\mu}\rangle (\phi) = \pi_\mu(\phi) + \int\limits_{-\infty}^\phi\!\ud\varphi\! \int\limits_0^\infty\!\ud\theta\; \mathcal{F}_\mu(\phi,\varphi,\theta) \;,
\ee
in which $\mathcal{F}_\mu$ is given explicitly in the Supplementary~\cite{Supp}; the details are not needed here. The important argument is $\theta$, which is the difference between phases at which photon emission occurs in the quantum state of the radiating system, and its complex conjugate. The $\theta$-integral contains quantum interference effects and is purely quantum mechanical, as it is confirmed by considering the low energy limit $b_0\ll 1$. In this limit the integrand collapses to a delta function in $\theta$~\cite{Ilderton:2013tb}, exhibiting decoherence~\cite{Zurek:2003zz} and leading to a purely local expression in agreement with classical predictions~\cite{Krivitsky:1991vt,Higuchi:2002qc,Higuchi:2004pr,Ilderton:2013tb,Moniz:1976kr}. Importantly, the classical limit of (\ref{P}) is closely related to the high-intensity limit. For high intensity (made precise below) the $\theta$-integrand is dominated by small perturbations around the classical point $\theta\simeq 0$. These semiclassical contributions give the LCA to $\langle \hat{P}\rangle$ at high-intensity. By analysing the momentum for arbitrary pulse shapes we show in the Supplementary~\cite{Supp} that the high-intensity and classical regimes are collectively characterised by the restriction
\be\label{boom}
	\frac{1+a_0^2}{b_0} \gg 1 \;.
\ee
This gives a kinematic refinement of the usual statement that only $a_0\gg1$ is required for the LCA to hold~\cite{Nynot}.  (The regime $a_0>1$ and $a_0^2> b_0$ has also been identified as that of the ``quantum synchrotron approximation''~\cite{KHKH}.) For other refinements coming from consideration of the emitted photon spectrum see~\cite{Khokonov:2002cf}. For ultra-intense optical lasers and achievable electron energies, (\ref{boom}) clearly implies $a_0\gg1$, but if either the energy is high or if the intensity is not so high so that (\ref{boom}) is {\it not} satisfied, quantum RR must be described using the full coherent expression~(\ref{P}). This integral contains correlations and interference between scattering events separated by arbitrarily large phase differences; it is this {\it quantum interference} which reduces RR energy losses as compared to the classical theory. Hence both the classical theory, which misses all interference effects, and the LCA, which captures only `short range' interference effects but misses the long range effects, overestimate RR losses.

We can now explain the behaviours seen in Fig.~\ref{FIG:SLAC}. The inequality (\ref{boom}) is not fulfilled:~because $b_0$ is not small enough and $a_0$ is not large enough, neither a low energy (local) nor a high-intensity (locally constant) approximation is valid. Rather the quantum interference effects in the coherent double-integral in (\ref{P}) are needed to properly capture QRR; when this holds we say that we are in an ``interference dominated regime'' (IDR). The simulation results in Fig.~\ref{FIG:SLAC} naturally overestimate the energy loss as they are based on the LCA, which misses quantum interference. This is consistent with recent investigations which show that the LCA misses spectral features which depend on long distance phase correlations or interference from multiple stationary points,  in both photon emission~\cite{Harvey:2014qla,Seipt:2015rda} and pair production~\cite{Meuren:2015mra,Nousch:2015pja,Jansen:2015idl}.

\begin{figure}[t!]
\includegraphics[width=0.9\columnwidth]{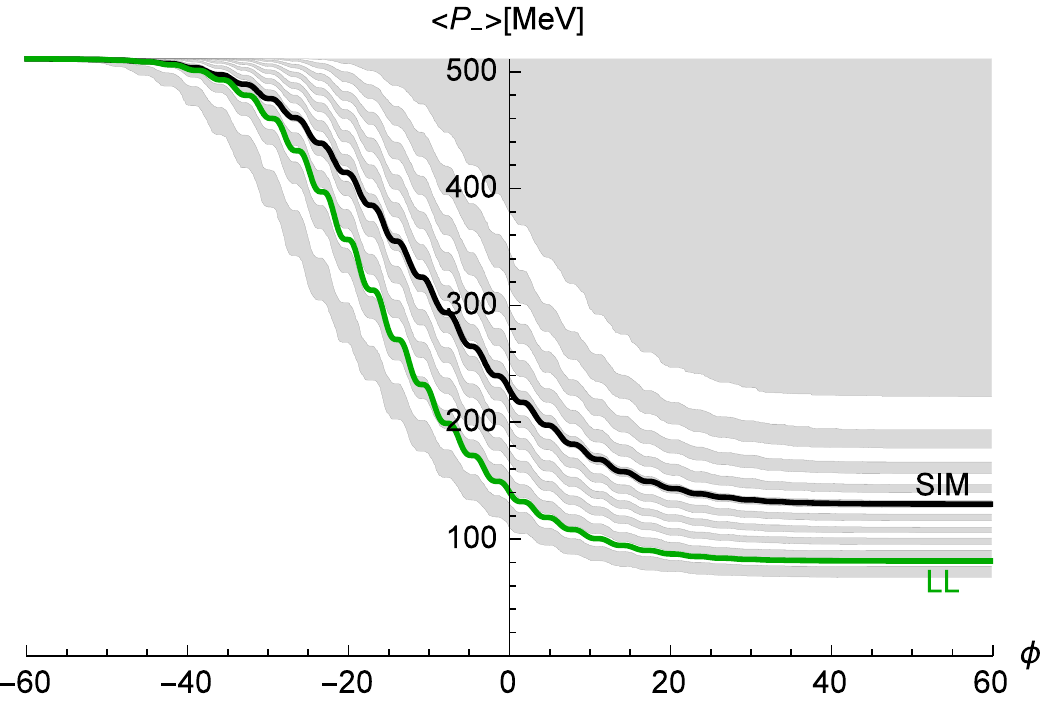}
\caption{\label{FIG:ELI-PW} $P_\LCm$ for $a_0=100$, $\gamma=10^3$ (high-intensity regime). The relative error in the classical energy loss (green) compared to the quantum multiphoton description (black) is only $15\%$. Grey/white bands illustrate the momentum distribution calculated with the numerical approach: each band contains, from top to bottom, $5\%$, $10\%$, $15\%\ldots$ of all trajectories.}
\end{figure}

We will now compare and contrast the IDR with the high-intensity regime accessible by the ELI-NP facility~\cite{ELI}. Taking $a_0 = 100$ and $\gamma = 10^3$ gives the same~$\chi$ as above, but in a different regime where (\ref{boom}) is satisfied. The LCA should therefore provide a good approximation here, in a regime where interference effects are suppressed and QRR comes from multiple incoherent photon emission~\cite{DiPiazza:2010mv}, and the numerical approach is on firm ground. (Entering an IDR for $a_0\gg1$ would, from (\ref{boom}) require extremely high energy particles.) Results are shown in Fig.~\ref{FIG:ELI-PW}. An average of 28.45 photons were emitted over $10^4$ simulation runs; higher-order multiphoton effects are indeed important. For this reason the order-$\alpha$ QED result is insufficient to capture the correct physics, and therefore not shown. Fig.~\ref{FIG:ELI-PW} shows that the difference between quantum and classical results is not large; the relative error in the classical prediction (an overestimate) is around~$15\%$, compared with around $350\%$ in the IDR. The reason for this is the high intensity; the system is driven back toward the classical regime as particles are shaken violently by the laser and very quickly radiate away their initial energy, well before reaching the peak field. The maximum~$\chi$ achieved is (from simulation data)~$\chi\simeq 0.25$, despite the initial parameters giving us a theoretical maximum $\chi \simeq 0.59$.  This resistance to entering the high-intensity, high-energy regime is well known~\cite{Pom,Fedotov} and is responsible for e.g.~hindering comparisons of different classical RR models~\cite{Kravets:2013}. In all our high $a_0$ simulations the number $N$ of photons emitted per laser cycle is consistent with the estimate, derived assuming a formation length $\sim 1/(\omega a_0)$~\cite{RitusReview}, $N\sim \sqrt{2}\pi \alpha a_0$ ($2\pi \alpha a_0$) for linear (circular) polarisation which differs from the commonly used~$N\sim\alpha a_0$.

\begin{figure}[t!]
	\includegraphics[width=0.9\columnwidth]{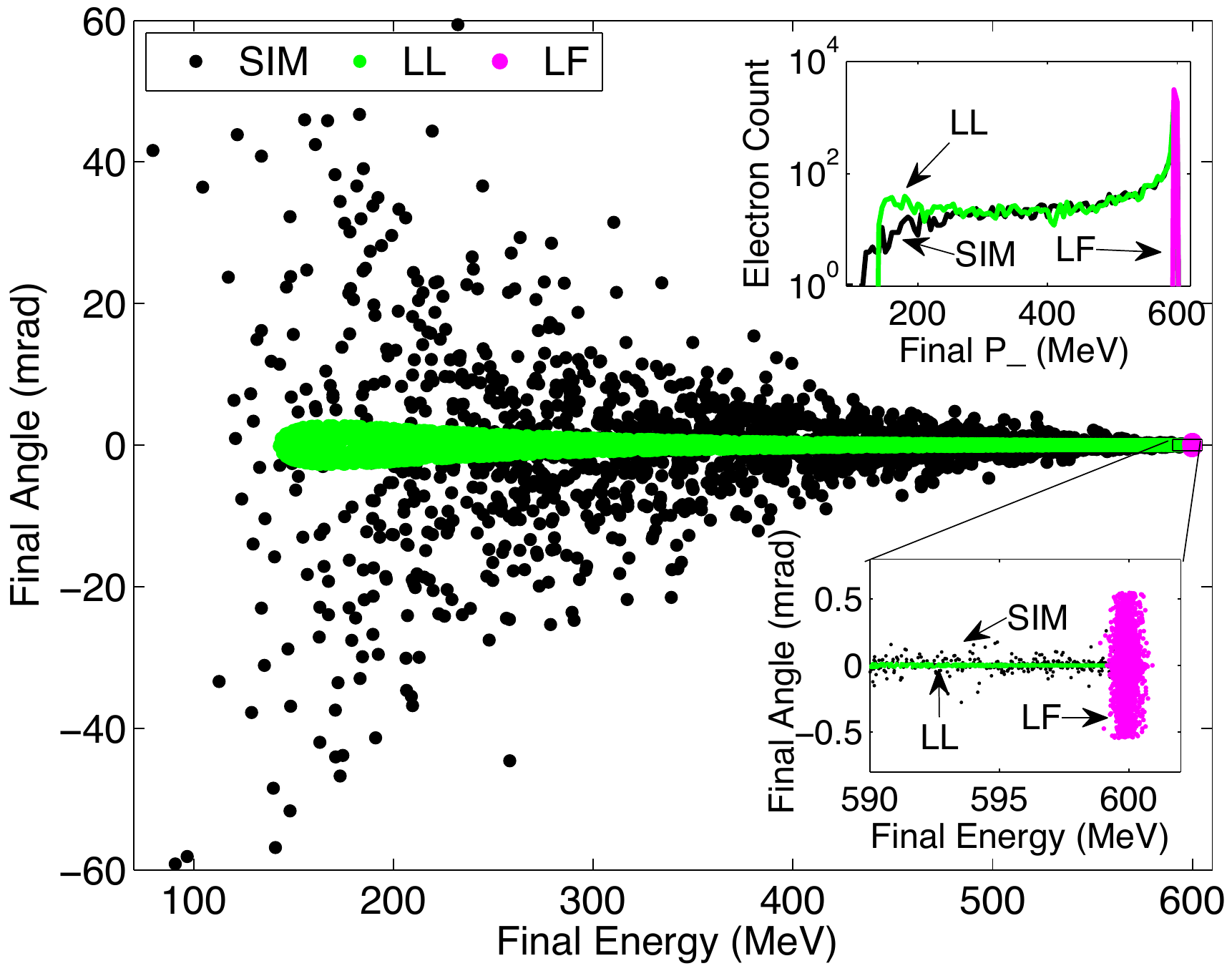}
	\caption{\label{FIG:ELI} Scattered electron spectrum for ELI-NP parameters, as in the text. We look along the energy axis to distinguish between classical models with and without RR, and along the angular scattering axis to distinguish between classical and quantum RR.}
\end{figure}

We turn now to two specific experimental scenarios in which signals of QRR will be sought in different regimes. The first extends the calculation above to a fully realistic collision of an electron beam with a focussed laser pulse, taking account of longitudinal and transverse beam structures and using the planned parameters of ELI-NP. We simulated a bunch of 5000 electrons with average energy $600\text{ MeV}$ ($\gamma\simeq1200$) $\pm0.1\%$ and transverse/longitudinal spread of FWHM $15\mu$m/400pm colliding with a focussed Gaussian pulse of wavelength $\lambda=820\text{ nm}$, focal spot radius $w_0 = 5\mu\text{m}$, FWHM pulse length $22 \text{ fs}$ and peak intensity $10^{22}$W/cm$^2$ ($a_0\simeq70$). The beam profiles are shown in the Supplementary~\cite{Supp}. Three simulations were performed, in which recoil effects were either neglected entirely (motion described only by the Lorentz force), treated classically (motion described by the Landau Lifshitz equation) or treated quantum mechanically using the numerical approach~\cite{Elkina:2010up,Ridgers,Green:2014kfa,Gonoskov:2014mda}. The results in Fig.~\ref{FIG:ELI} show marked differences between the three models.

Looking along the energy axis shows that both classical and quantum RR cause the electron beam to emerge from the pulse with an energy spread of several hundreds of MeV, whereas neglecting recoil effects implies that the electrons essentially retain their initial energies~\cite{Neitz,Yoffe}. However distinguishing quantum and classical contributions to this effect is difficult, see also the top inset of Fig.~\ref{FIG:ELI}. This is because beam focussing (finite width with varying intensity) gives an impact-parameter spread in energy which acts as a background. Looking instead along the vertical axis, corresponding to transverse scattering angle, we see that the quantum electrons develop a transverse spread spanning several degrees, corresponding to a transverse momentum spread of around~$10$MeV, whereas the classically modelled electrons remain largely confined to the beam axis.  The stochastic nature of quantum emissions means that the electron bunch will diffuse in transverse momentum space~\cite{Green:2013sla}, whereas the classical model only allows a net {\it loss} of transverse momentum in the radiating electrons, with the exception of comparatively much smaller ponderomotive effects (a small background) arising from beam focussing. Hence the transverse spreading provides a measurable signature of quantum RR distinct from its classical counterpart. (Transverse size effects are expected to be subleading in the IDR with high energy particles: the highest energy emissions come from particles on-beam-axis~\cite{Mackenroth:2010jk} and transverse deflection from the plane wave trajectory is suppressed by factors of $a_0/\gamma \ll 1$~\cite{DiPiazza:2013vra}.)

Finally, consider the regime $a_0\sim 1$--$10$ and $\gamma\leq 10^4$ which should be accessible on the Bella~\cite{Bella} and Gemini~\cite{Gemini,GLS} lasers. Toward the lower/upper extreme of the energy range classical/quantum effects are significant. Toward the lower/upper extreme of the intensity range the LCA fails/works, so that simulations are less/more reliable. At the same time the lowest order QED results become more/less reliable, because the longer or more intense the pulse, the more higher-order corrections are required to account for multiple photon emissions in order to give the correct rate of energy loss. In this ``crossover'' regime it is therefore necessary to account carefully for the possibility of both multiphoton and interference effects. This regime is distinct from the IDR and high-intensity regimes above. It is theoretically challenging, as higher-order corrections are difficult to calculate analytically when the constraint~(\ref{boom}) is not fulfilled~\cite{Ilderton:2013tb}.
\begin{figure}[t!]
	\includegraphics[width=0.95\columnwidth]{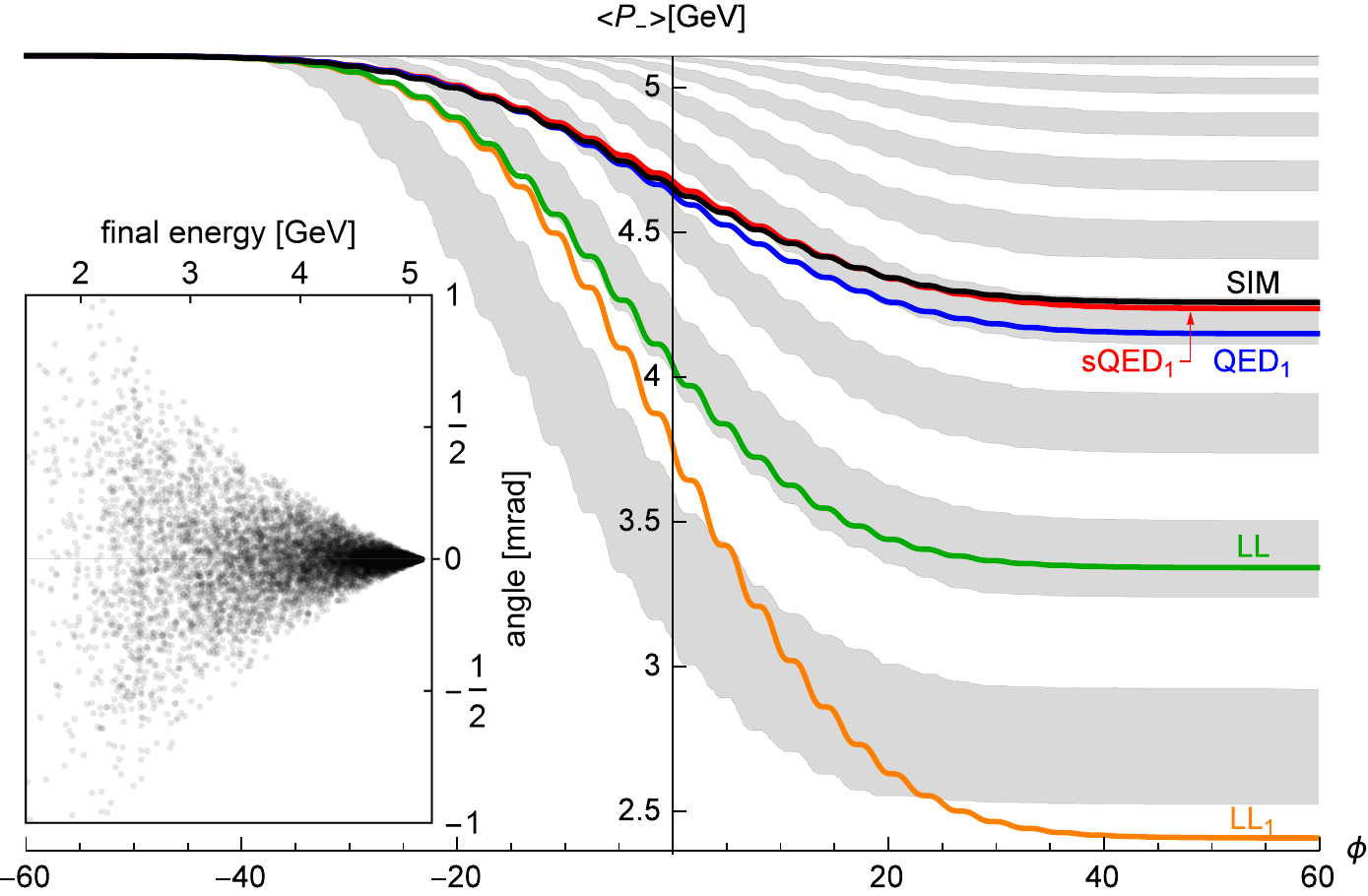}
	\caption{\label{FIG:GEMINI} $\langle P_\LCm \rangle$ for a head-on collision in the crossover regime, colours as above.  $a_0=10$ and $\gamma=10^4$. Just over $60\%$ of trajectories show lower-than-average energy loss. The inset shows the final distribution of electron energies vs.~scattering angle, due to stochastic quantum effects.}
\end{figure}

An example of the electron momentum in the crossover regime is shown in Fig.~\ref{FIG:GEMINI} for parameters giving the same~$\chi$ as above. Here the LCA is sufficient to capture the physics -- the LCA to (\ref{P}) is indistinguishable from the full result on the scale shown. Since an average of 2.77 photons were emitted over $10^4$ runs we should expect a discrepancy between the simulation and order-$\alpha$ QED results due to multiphoton effects. Despite this, we find that they are in close mutual agreement (and both differ significantly from the classical prediction). Remarkably, the same agreement is found for all other parameters we have examined in this regime. The fact that the two very different approaches agree across an energy and intensity range relevant to upcoming experiments, e.g.~on Gemini, is extremely encouraging. We stress though that further investigation of this interesting regime is needed to ensure that the correct result is obtained.

One reason for the smallness of the expected discrepancy in Fig.~\ref{FIG:GEMINI} can be seen by from the grey bands: most electron trajectories stay distributed close to the Lorentz-force trajectory for most of the pulse. The distribution and spread of momenta is therefore an interesting topic for further study, along with the quantum mechanical variance $\langle \hat{P}^2 \rangle - \langle \hat{P}\rangle^2$.The inset in Fig.~\ref{FIG:GEMINI}, shows the stochastic spreading of on-axis electrons due to purely quantum effects, c.f.~Fig.~\ref{FIG:ELI}. This is one of the experimental signatures of QRR which will be investigated at high-power laser facilities over the coming years.

\begin{figure}[t!]
	\includegraphics[width=0.95\columnwidth]{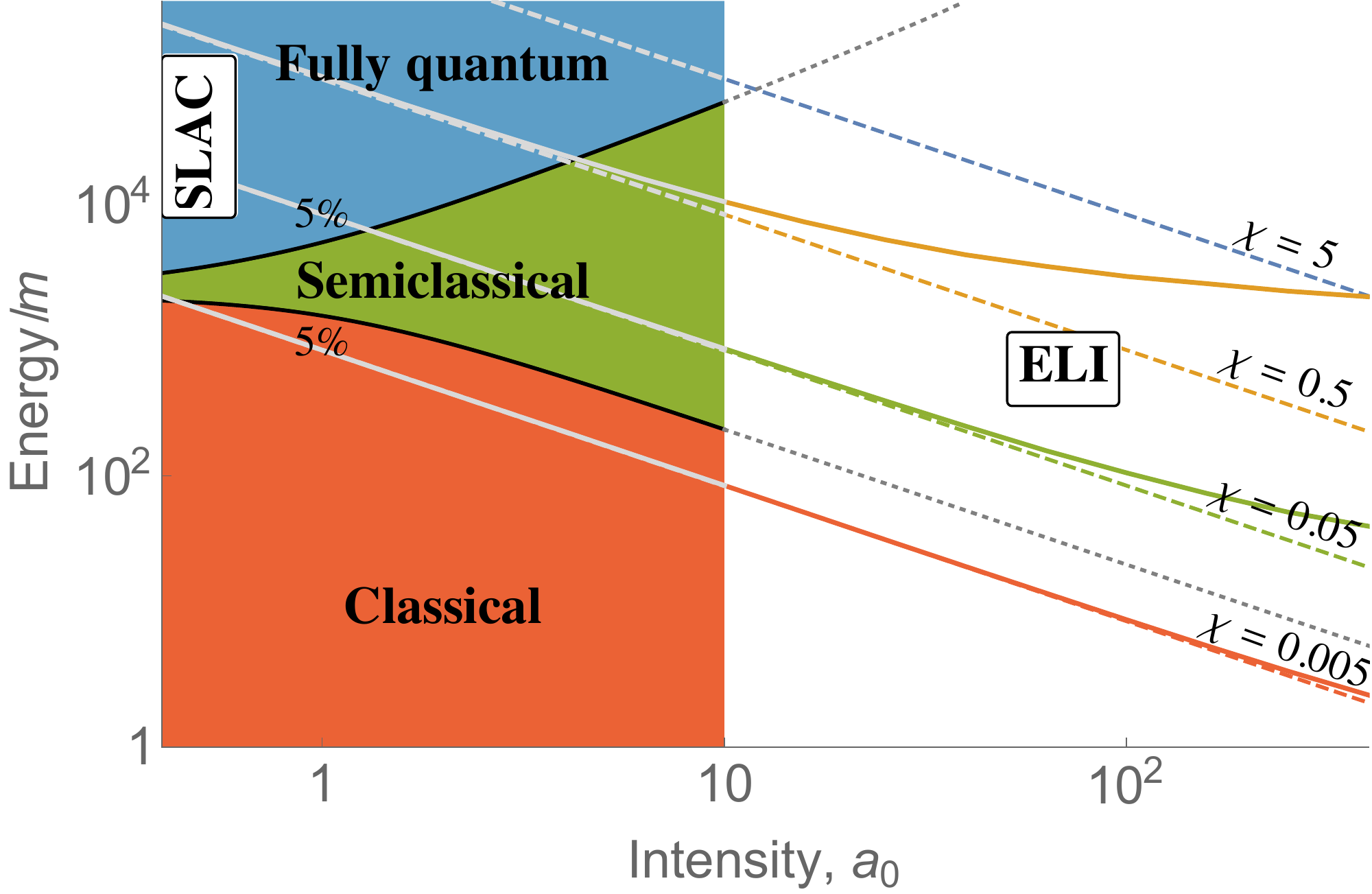}
	\caption{\label{FIG:TECKNING} Characterising radiation reaction in the energy--intensity plane. Two types of effects are shown, to the left and right of the division at $a_0=10$ which separates, very approximately, the applicability of the methods used. The classical (semiclassical/locally constant) approximation to RR differs by less than 5\% from the full, coherent, quantum integral (\ref{P}) in the regions marked ``classical'' (``semiclassical'') on the left of the plot. As intensity increases, higher energies are needed to access fully quantum effects. To the right, dotted lines are those of constant ``ideal'' $\chi$ calculated from peak intensity and initial energy, while solid lines are those of constant peak $\chi$ taking into account classical cooling effects modelled by the LL equation. The existence of the different regimes and cooling effects are general, though their precise form depends on pulse shape. Labels in white boxes indicate the approximate operating regimes of the named facilities~\cite{Bamber:1999zt,ELI}.}
\end{figure}

To conclude, we have examined QRR effects in different energy and intensity regimes. Fig.~\ref{FIG:TECKNING} illustrates these regimes and our results.  We have seen that interference effects, completely absent in classical physics, reduce energy losses relative to classical predictions and contribute significantly to QRR for high energy and not too high intensity.  This ``interference dominated regime'', or IDR, stands in contrast to the high-intensity regime in which quantum RR is essentially semiclassical and captured by a locally constant approximation, and where large cooling effects draw systems back toward the classical regime~\cite{Mnote}. We have also identified a kinematic delineation of these different regimes, see also~\cite{KHKH,SVB}; refinements accounting for pulse duration~\cite{King:2013osa} or final state kinematics~\cite{Khokonov:2002cf} are interesting topics for future study.

Concerning experimental signatures, we have confirmed that both classical and quantum radiation reaction will be visible in high-intensity ELI-NP experiments. We have also highlighted a ``crossover regime'' where both multiphoton and quantum interference effects are significant. This is the most interesting, theoretically challenging, and perhaps experimentally urgent regime.

\textit{The authors are supported by a Strategic Grant POSDRU/159/1.5/S/137750 (V.D.), the Olle Engkvist Foundation, grant 2014/744 (A.I.), the Wallenberg Foundation project ``Plasma based compact ion sources'' (A.I., M.M.) and the Swedish Research Council, grants 2011-4221 (A.I., G.T), 2012-5644, 2013-4248 (C.H., M.M).}

\appendix
\newpage
\onecolumngrid
\vspace{20pt}
\begin{center}
	\textbf{\Large{Supplementary Material}}
\end{center}
\section{Analytic results and approximations}
A plane wave $F_{\mu\nu}(\phi)$ depends on a phase variable $\phi=k\cdot x$ with $k^2=0$ defining the propagation direction. The fields of the wave are transverse, $k^\mu F_{\mu\nu}=0$, and can be projected out by defining the vector $\bar{k}^\mu$ by $k\cdot\bar{k}=1$; the two nonzero field components are then $\bar{k}^\mu F_{\mu\nu}$. The Lorentz force equation, describing the motion of a classical particle but neglecting recoil effects, can be solved exactly in a plane wave and the resulting particle momentum is
\be
	\pi_\mu(\phi) = p_\mu - a_\mu(\phi) + \frac{2 p\cdot a(\phi) - a(\phi)^2}{2k.p} k_\mu \;,
\ee
in which $p_\mu$ is the (initial) momentum in the remote past, or before entering the wave, and the integrated field strength $a(\phi)$ is
\be
	a_\nu(\phi) = \int\limits_{-\infty}^\phi\!\ud\varphi\; e \bar{k}^\mu F_{\mu\nu}(\varphi) \;.
\ee
In QED the expectation value $\langle \hat{P}_\nu \rangle$ can also be solved for exactly in the background field strength and order-by-order in the fine structure constant $\alpha$. Beginning with a state $\ket{\Psi;0}$ describing an initial electron, the state is evolved in time to $\ket{\Psi;\phi}$ (following~\cite{XNeville:1971uc}
we use phase as the time variable, as explained in the Letter) and the expectation value is calculated as $\langle \hat{P}_\nu \rangle \equiv \bra{\Psi;\phi} \hat{P}_\nu\ket{\Psi;\phi}$. The method of calculation is described in detail in~\cite{XIlderton:2013dba} for scalar QED; spin corrections are represented by additional, similar terms in the expressions therein. 

\begin{figure}[b!]
	\includegraphics[width=0.5\textwidth]{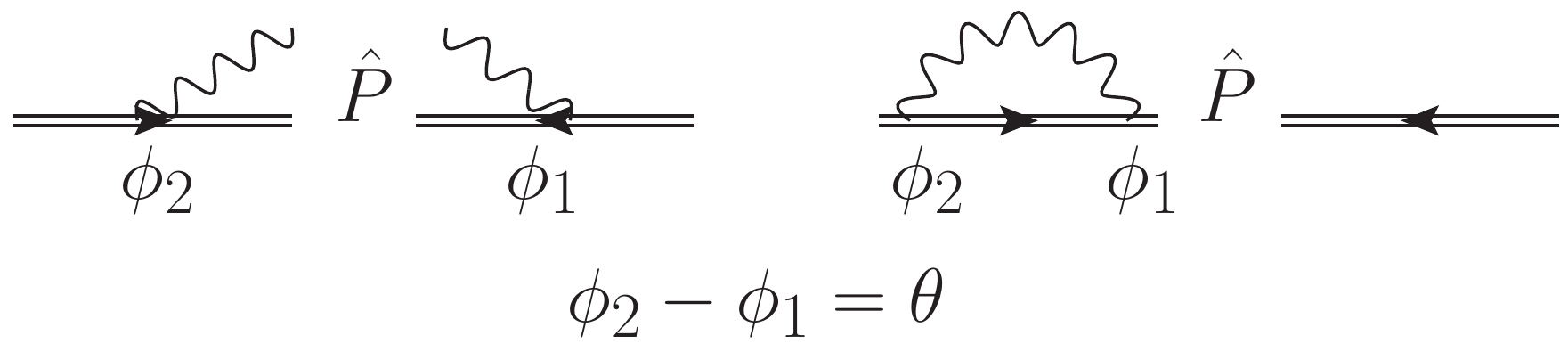}
	\caption{\label{FIG:THETA} Schematic illustration of some of the terms contributing to the expectation value $\langle {\hat P} \rangle$. The diagrams represent the state wavefunctional and its conjugate. Photon emission contributions are illustrated on the left, self energy contributions on the right. The argument $\theta$ arises as the phase-difference between points of  emission/absorption.}
\end{figure}

To first nontrivial order in $\alpha$ the average momentum naturally receives contributions from photon emission diagrams (at tree level), and also from one-loop self energy effects. The loop is essential for removing infra-red divergences and ensuring the existence of the classical limit~\cite{XHolstein:2004dn,XHiguchi:2004pr,XBrodsky:2010zk,XIlderton:2013dba}. To this order the momentum (averaged over spins) takes the form
%
%
\be\label{IJ}
	\langle {\hat P}_\nu \rangle(\phi)  = \pi_\nu(\phi) +\frac{\alpha}{\pi} \int\limits_{-\infty}^{\phi}\!\ud\phi_2 \int\limits_0^\infty\!\ud \theta\; \bigg[\frac{\ud \mathcal{I}_\nu}{\ud\phi_2} + \frac{1}{b_0} \mathcal{J}_\nu\bigg]  \;,
\ee
in which $\phi_2$ and $\theta$ are two phases originating from the two interaction vertices in the expectation value. Note that  $\theta$ is the phase difference between the vertices, see Fig.~\ref{FIG:THETA}. $\mathcal{I}$ and $\mathcal{J}$ are functions depending on the integration variables, the particle spin, initial momentum and the background field. Note that the first term is an exact integral which is to be evaluated on the upper boundary $\phi_2=\phi$. 

The explicit expressions below hold for the lower $\{-,\perp\}$ components of the momentum. For the $+$ component there are additional, similar terms, all proportional to $k_\nu / b_0$ and (therefore) carrying an extra factor of electron gamma in the denominator. These terms are small (in the lab frame) for both head-on and $45^\circ$ collisions with large gamma factor (which we assume throughout, as otherwise RR is negligible in the short pulses considered -- for long pulses see though~\cite{XHeinzl:2015eda}). There are many such terms, but their expressions are lengthy and unrevealing. For these reasons we do not display the $k_\nu/b_0$ terms.

The expressions below for $\mathcal{I}$ and $\mathcal{J}$ use the following notation. Setting ${\bf s}=0/1$ switches between scalar/spinor QED respectively. The Lorentz force momentum is $\pi_\nu(\phi)$ and we write $\pi_{2\nu}\equiv \pi_\nu(\phi_2)$, $\pi_{1\nu}\equiv \pi_\nu(\phi_2-\theta)$. The average~$\langle \pi_\nu \rangle$ which appears is defined as
\be
	\langle \pi_\nu \rangle =\frac{1}{\theta}\int\limits_{\phi_2-\theta}^{\phi_2}\!\ud\varphi\; \pi_\nu(\varphi)
\ee 
Define $\mu \equiv M^2/m^2$, with the effective mass given as usual by $M^2 \equiv  \langle \pi\rangle^2$~\cite{XKibble:1975vz,XHebenstreit:2010cc}, and $\Theta \equiv \theta \mu$. Then we have:
%
\be
	\mathcal{I}_\nu =
	\frac{\pi_\nu\frac{\partial_\theta\Theta}{\mu}-\pi_{1\nu}}{\theta} \operatorname{Re} f_0 \bigg(  \frac{\Theta}{2b_0} \bigg) 
	 + \frac{  \pi_{1\nu} + \langle \pi_\nu\rangle   -  2\pi_\nu\frac{\partial_\theta\Theta}{\mu}}{2\theta} \operatorname{Re}f_{1} \bigg(  \frac{\Theta}{2b_{0}}\bigg)  + {\bf s} \frac{\frac{\partial_{\theta}\Theta
}{\mu}\pi_{\nu}-\left\langle \pi_{\nu}\right\rangle }{2\theta}  \operatorname{Re}f_{2}\bigg(\frac{\Theta}{2b_{0}}\bigg) 
\ee

\be
\mathcal{J}_\nu = \left[  \frac{ \frac{\partial_\theta\Theta}{\mu} \pi_{2\nu}-\left\langle \pi_{\nu}\right\rangle }{\theta}-\frac{\pi_{2\nu}^{\prime}}{2}+\frac{\left\langle \pi^{\prime}\right\rangle ^{2 }\theta}{2 m^{2}}\left\langle \pi_{\nu}\right\rangle \right]  \operatorname{Im}f_{1}\left(  \frac{\Theta
}{2b_{0}}\right) + {\bf s} \frac{\left\langle \pi^{\prime}\right\rangle ^{2}%
\theta}{4 m^{2}} \left\langle \pi_{\nu}\right\rangle \operatorname{Im}g_{2}\left(
\frac{\mu\theta}{2b_{0}}\right) 
\ee
The functions $f$ and $g$ arise in the calculation as integrals over the momentum fraction $u\equiv k\cdot k'/k\cdot p$, that is the ratio of the emitted photon's (longitudinal) momentum $k'$ to the final electron (longitudinal) momentum:
\be\label{fg2}
	 f_n(x) = \int\limits_0^1\!\ud u\; u^n e^{i \frac{u}{1-u} x} \;, \qquad g_n(x) = \int\limits_0^1\!\ud u\; \frac{u^{n+1}}{1-u} e^{i \frac{u}{1-u} x} \;.
\ee
We note that the presence of recoil effects are signalled in part by the presence of the $1/(1-u)$ factors in the exponent~\cite{XBK}. Expanding these factors $1/(1-u) \to 1$ amounts to ignoring corrections from large emitted photon energies (and therefore significant recoil), and to taking the low energy limit, see~\cite{XKhokonov:2002cf,XHarvey:2009ry} for discussions. We stress though that no such approximation is made here. Indeed the integrals (\ref{fg2}) may be evaluated exactly in terms of sine and cosine integrals. The explicit expressions are unrevealing, though. Changing variables $t\equiv u/(1-u)$ instead gives a more convenient, and explicitly convergent, form:
\be\label{fg1}
	f_n(x) = \int\limits_0^\infty\!\ud t \frac{t^n}{(1+t)^{n+2}}e^{it x} \;, \qquad g_n(x) = -i f'_n(x) \;.
\ee
\begin{figure}
	\includegraphics[width=0.5\columnwidth]{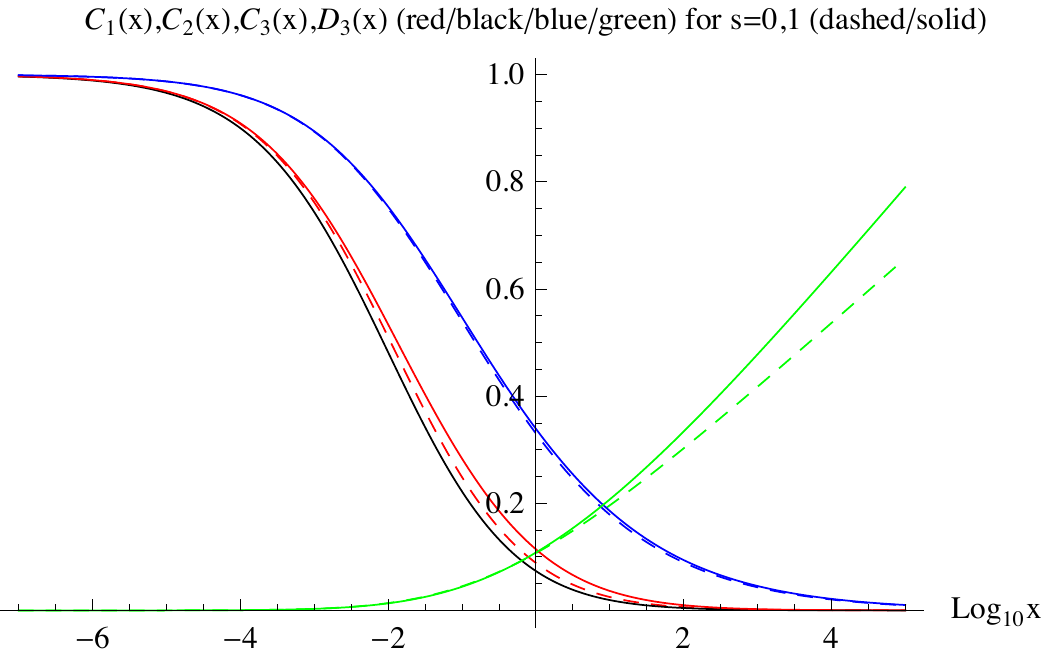}
	\caption{\label{FIG:CD} Behaviour of the special functions relevant to the high-intensity/locally constant approximation.}
\end{figure}
Our expressions for the average momentum can be put into an equivalent, but perhaps more familiar form in the asymptotic limit $\phi\to\infty$, where we find perfect agreement with $S$-matrix based calculations~\cite{XIlderton:2013tb}. Let $\bar\pi$ be the Lorentz force (i.e.~no recoil) prediction for the electron momentum after it leaves the pulse (typically equal to the initial momentum $p$.). Then the expectation value can be written, schematically,
\be
	\langle P_\nu \rangle = \bar{\pi}_\nu + \int\!\ud k'\; \big(p^{nlc}_\nu -  \bar{\pi}_\nu\big) \frac{\ud \mathbb{P}^{nlc}}{\ud k'} \;,
\ee
in which $\ud \mathbb{P}^{nlc} / \ud k'$ is the differential probability for nonlinear Compton scattering, i.e.~for the electron to emit a photon of momentum $k'$~\cite{XNikishov:1963}, and $p^{nlc}$ is the electron momentum after emission as dictated by momentum conservation~\cite{XNikishov:1963}. For precise expressions see equations (5), (8) and (9) in~\cite{XIlderton:2013tb}. Thus our results admit a very natural interpretation: the electron momentum changes because it emits radiation, just as we expect, and the resulting average momentum is found by averaging over all possible emissions, weighted with the probability of photon emission. The probability here agrees with that in~\cite{XNikishov:1963,XHarvey:2009ry} for the case of a monochromatic wave, with those of~\cite{XRitusReview} in the locally constant field approximation~\cite{XHarvey:2014qla}, and with those in e.g.~\cite{XBoca:2009zz,XDinu:2012tj,XSeipt:2014yga} in the case of a pulse.

\subsection{Classical and local approximations}
Note that the argument of the special functions, $\theta \mu/2b_0$, is an increasing function of $\theta$. If this function becomes large outside a very small vicinity of $\theta=0$ the functions $f$ and $g$ become negligibly small. In this case we can thus develop a {\it local} approximation to the general formula (\ref{IJ}). This happens when, compared to unity, $b_{0}$ is small, $a_{0}^{2}$ is large (due to the fast increase of the effective mass with $\theta$), or both. In other words, using the form of the effective mass, when $(1+a_0^2)/(2 b_0) \gg 1$. In this regime quantum radiation reaction becomes approximately incoherent, arising from only the local behaviour of the field, with no interference between emission at different phases. A Taylor expansion in $\theta$ then leads to the general local approximation (again up to $k/b_0$ terms)
\be\label{LCAfull}
	\langle {\hat P} \rangle_\nu = 
	\pi_\nu + \alpha \bigg[ b_0 \pi_\nu^\prime\mathcal{C}_3(\lambda) 
	+ \pi_\nu \mathcal{D}_{3} (\lambda)  
	+ \frac{2}{3}b_{0}\int_{-\infty}^{\phi}\!\ud\phi_2 a_2^{\prime2} \pi_{2\nu}\mathcal{C}_{1}(\lambda_{2}) 
	-\frac{\pi_{2\nu}^{\prime\prime}}{2}\mathcal{C}_2 (\lambda_2)  \bigg]
\ee
where $\lambda=-b_0^2 a^{\prime 2}(\phi)/3$ and $\lambda_2 = -b_0^2 a^{\prime 2}(\phi_2)/3$.  The $\mathcal{C}_i$ and $\mathcal{D}_i$ are special functions which we do not give explicitly but which are plotted in Fig.~\ref{FIG:CD}. The behaviour of the functions $\mathcal{C}_1$ in the scalar and spinor cases shows directly that spin increases the effect of RR, i.e.~leads to larger RR losses.

Fig.~\ref{FIG:CD} shows that the convergence of the special functions towards their limit at the origin ($\mathcal{C}_{i}(0)  =1$ and $\mathcal{D}_{i}(0)  =0$) is not so rapid, implying that a significant deviation from the local approximation can be expected even for $\chi$ less than $10^{-2}$. At even smaller $\chi$ we recover classical results: the local approximation reduces to, reinstating the $k_\nu/b_0$ terms to demonstrate that the momentum goes on-shell,
\be\label{LCAclassical}
	\langle {\hat P} \rangle_{\nu}=\pi_{\nu}+\frac{2}{3}\alpha b_{0} \pi_{\nu}^{\prime} + \frac{2}{3}\alpha b_{0}  \int\limits_{-\infty}^\phi\!\ud\phi_2\;  (a_2^\prime)^2 \bigg(  \pi_{2\nu} -\pi\cdot\pi_2 \frac{k_\nu}{b_0} \bigg)   \;,
\ee
which is of course the first order perturbative solution to the LAD (and LL) equations. The complete local approximation (\ref{LCAfull}) is useful in the regime of moderately large $a_{0}$ and moderately small $b_{0}$. It can be further simplified in both the classical limit, (\ref{LCAclassical}), and in the high intensity limit $a_{0}\gg1$, where it becomes
\be\label{LCCC}
	\langle {\hat P} \rangle_\nu \simeq \pi_\nu + \frac{2}{3}\alpha b_{0}\int_{-\infty}^{\phi}d\phi
_{2}\left(  a_{2}^{\prime}\right)  ^{2}  \pi_{2\nu}\mathcal{C}_1 (\lambda_2)  \;.
\ee
In the plots for the crossover regime, in the Letter and below, the result of the approximation (\ref{LCCC}) is indistinguishable from that of the coherent calculation (\ref{IJ}). \\[-5pt]
\newpage
\twocolumngrid
\section{Transverse structure}
\begin{figure}[h!!!]
	\includegraphics[width=0.9\columnwidth]{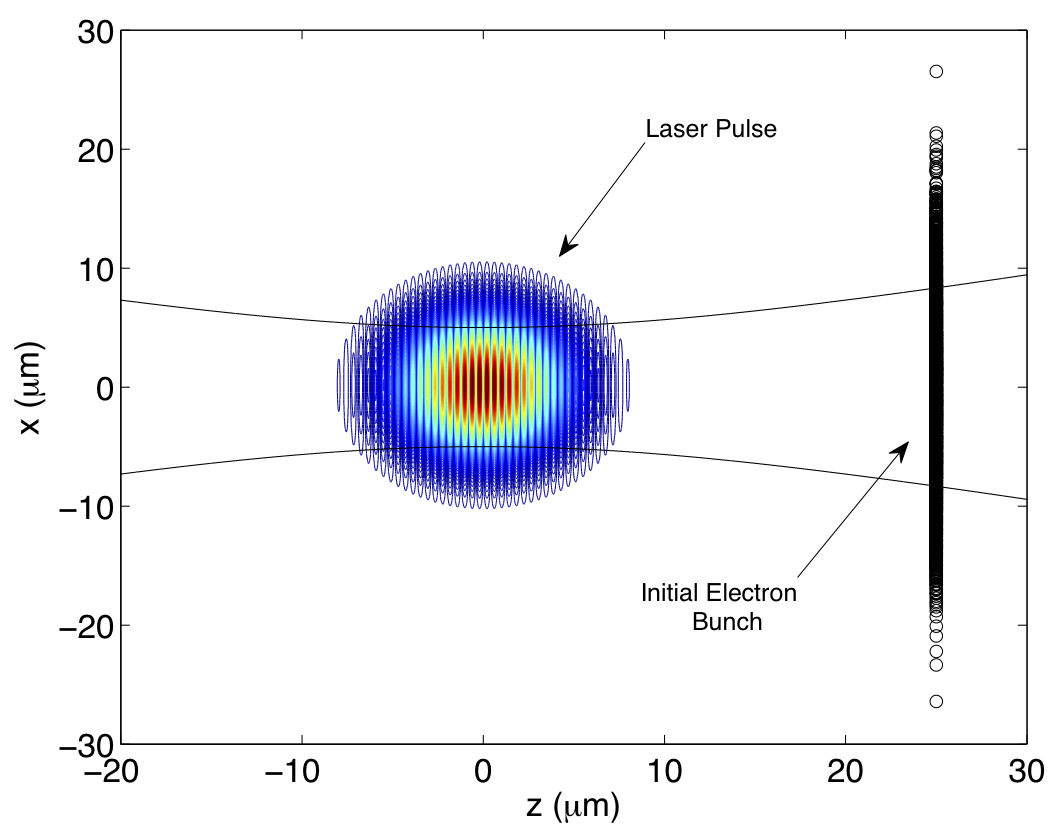}
	\caption{\label{FIG:GEMINI45} The paraxial Gaussian laser pulse and electron bunch used for simulation of experiments at ELI-NP, see Fig.~3 in the Letter.  \textit{Laser:} wavelength $\lambda=820\text{ nm}$, focal spot radius $w_0 = 5\mu\text{m}$, FWHM pulse length $22 \text{ fs}$ and peak intensity $10^{22}$W/cm$^2$ ($a_0\simeq70$). Solid black lines show how the laser waist size changes as the pulse propagates. \textit{Electron bunch:} 5000 electrons with average energy $600\text{ MeV}$ ($\gamma\simeq1200$) $\pm0.1\%$ and transverse/longitudinal spread of FWHM $15\mu$m/400pm.}
\end{figure}

\section{45$^\circ$ collisions}
\noindent Collision at $45^\circ$ incidence, as may be experimentally necessary, can be advantageous as it makes QRR visible in all momentum components, see the following examples.
\begin{figure}[h!]
	\includegraphics[width=\columnwidth]{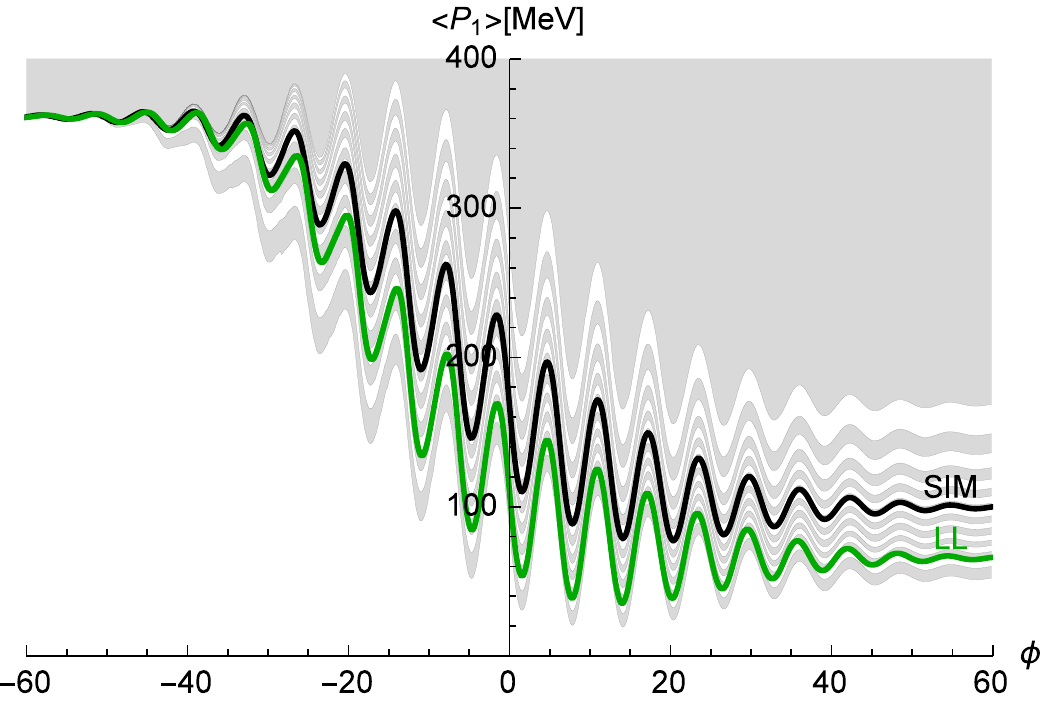}
	\caption{\label{FIG:ELI-45} Perpendicular electron momentum in the high-intensity regime, $a_0=100$ and $\gamma=10^3$, but for a $45^\circ$ collision. Grey and white bands show, top to bottom, intervals which contain $5\%$, $10\%$, $15\%\ldots$ of all trajectories calculated with the numerical approach.}
\end{figure}
\newpage
%
\begin{figure}[h!]
	\includegraphics[width=\columnwidth]{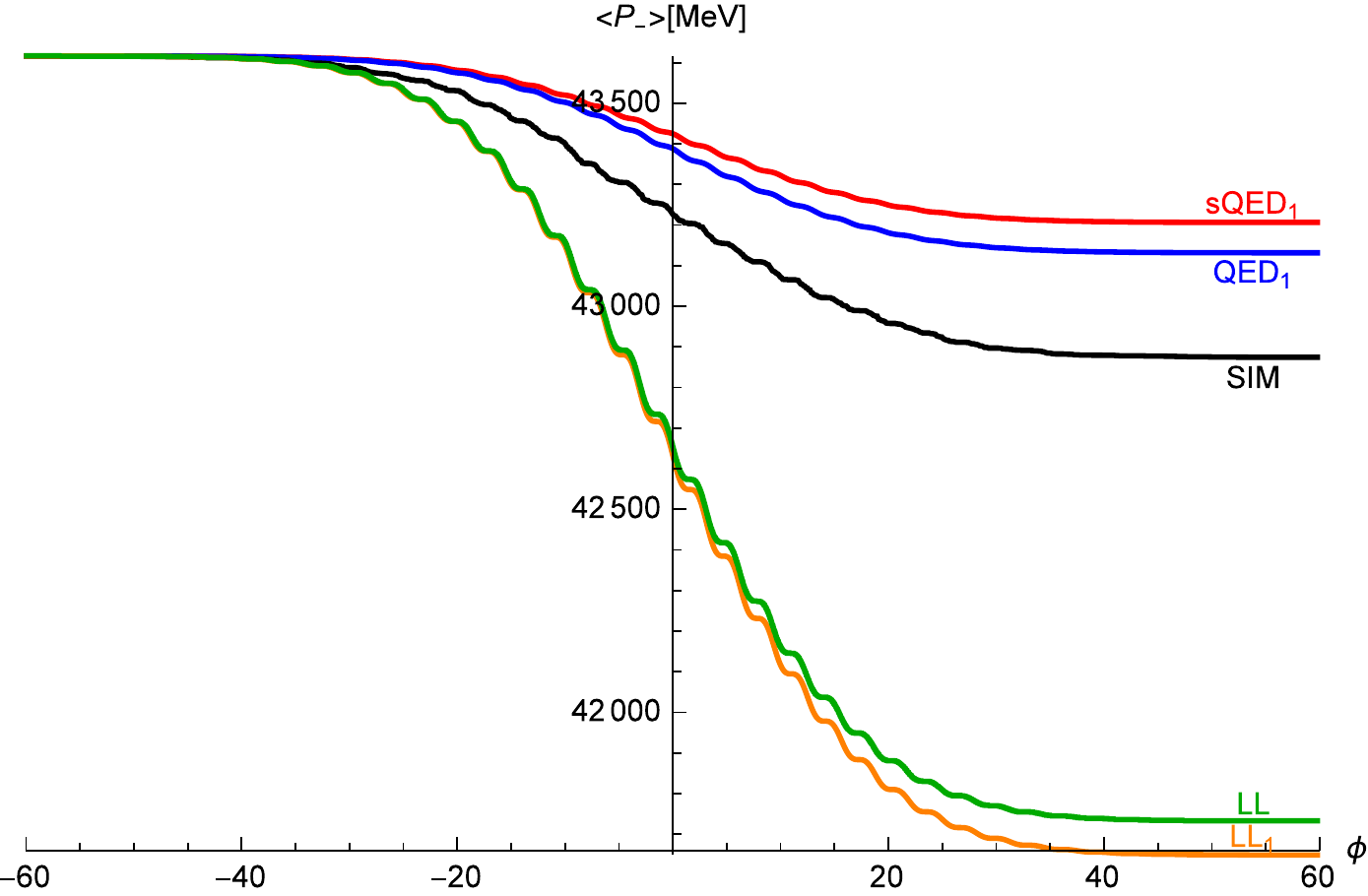}
	\includegraphics[width=\columnwidth]{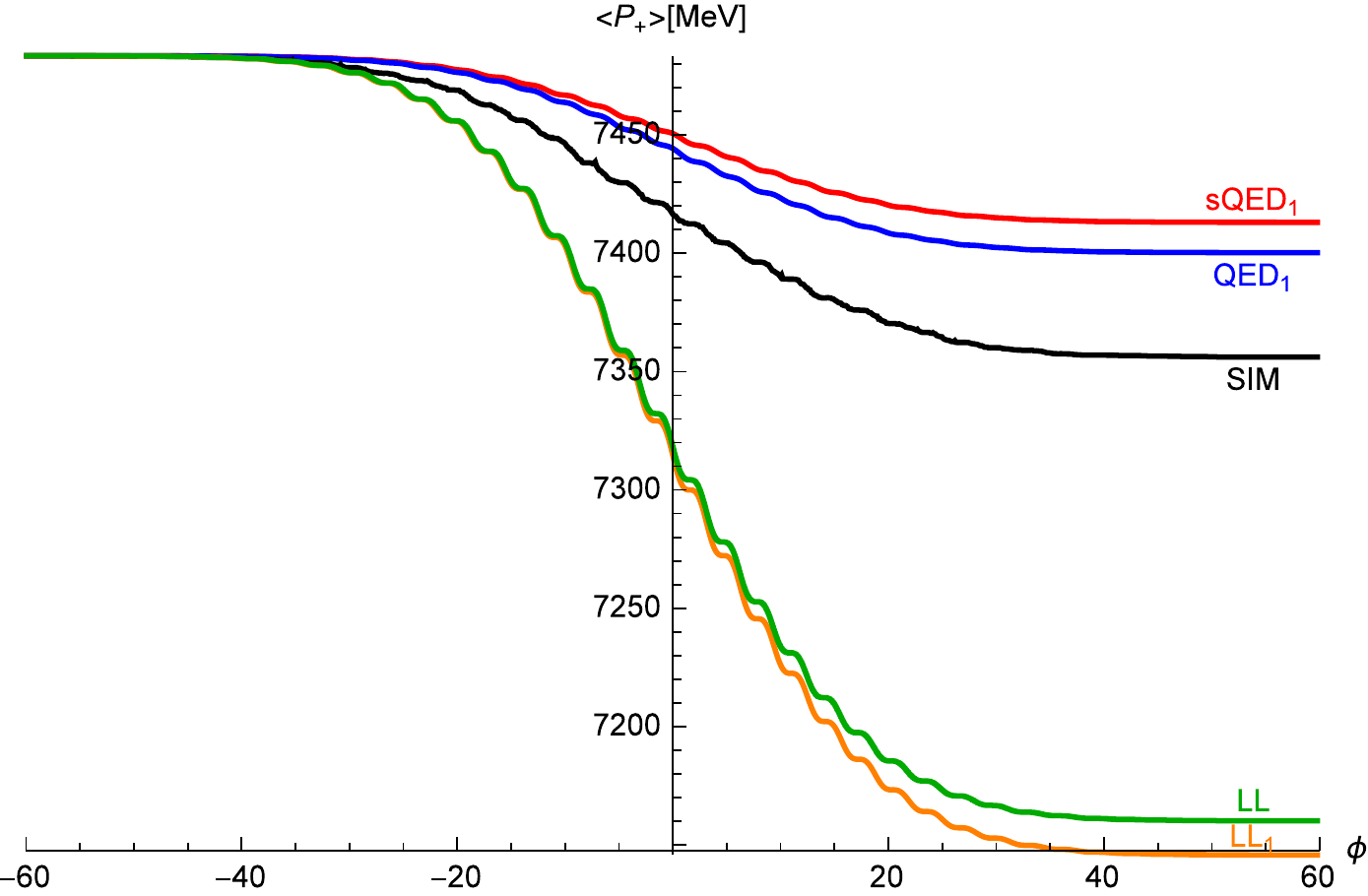}
	\includegraphics[width=\columnwidth]{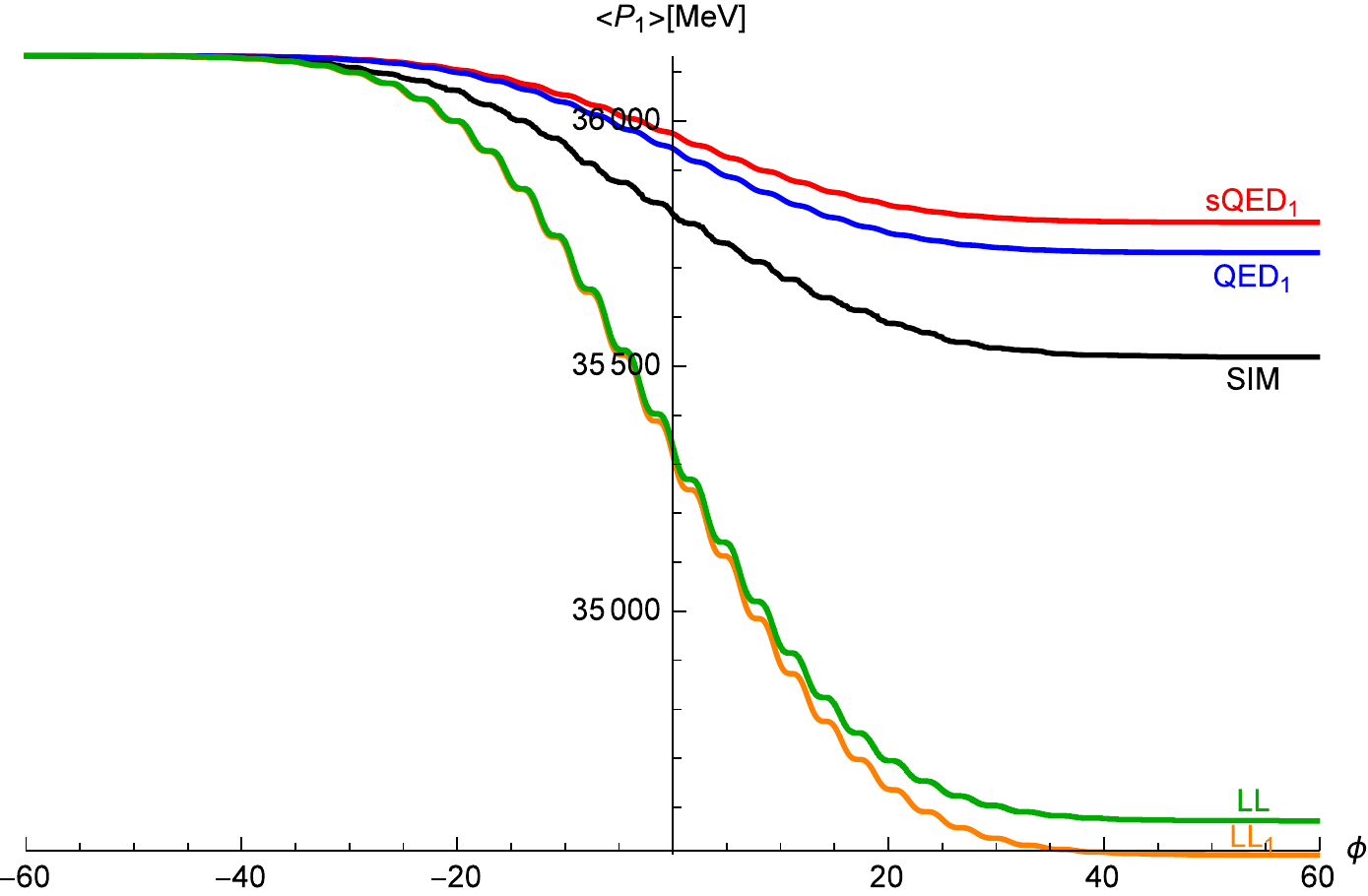}
	\caption{\label{FIG:SLAC-45} All nontrivial electron momentum components in the interference dominated regime, $a_0=1$ and $\gamma=10^5$, but for a $45^\circ$ collision. Colours as in the Letter. QRR effects of similar size are visible in all components. Classical predictions overestimate RR losses, as does the quantum but locally constant approximation.}
\end{figure}
\begin{figure}[h!]
	\includegraphics[width=\columnwidth]{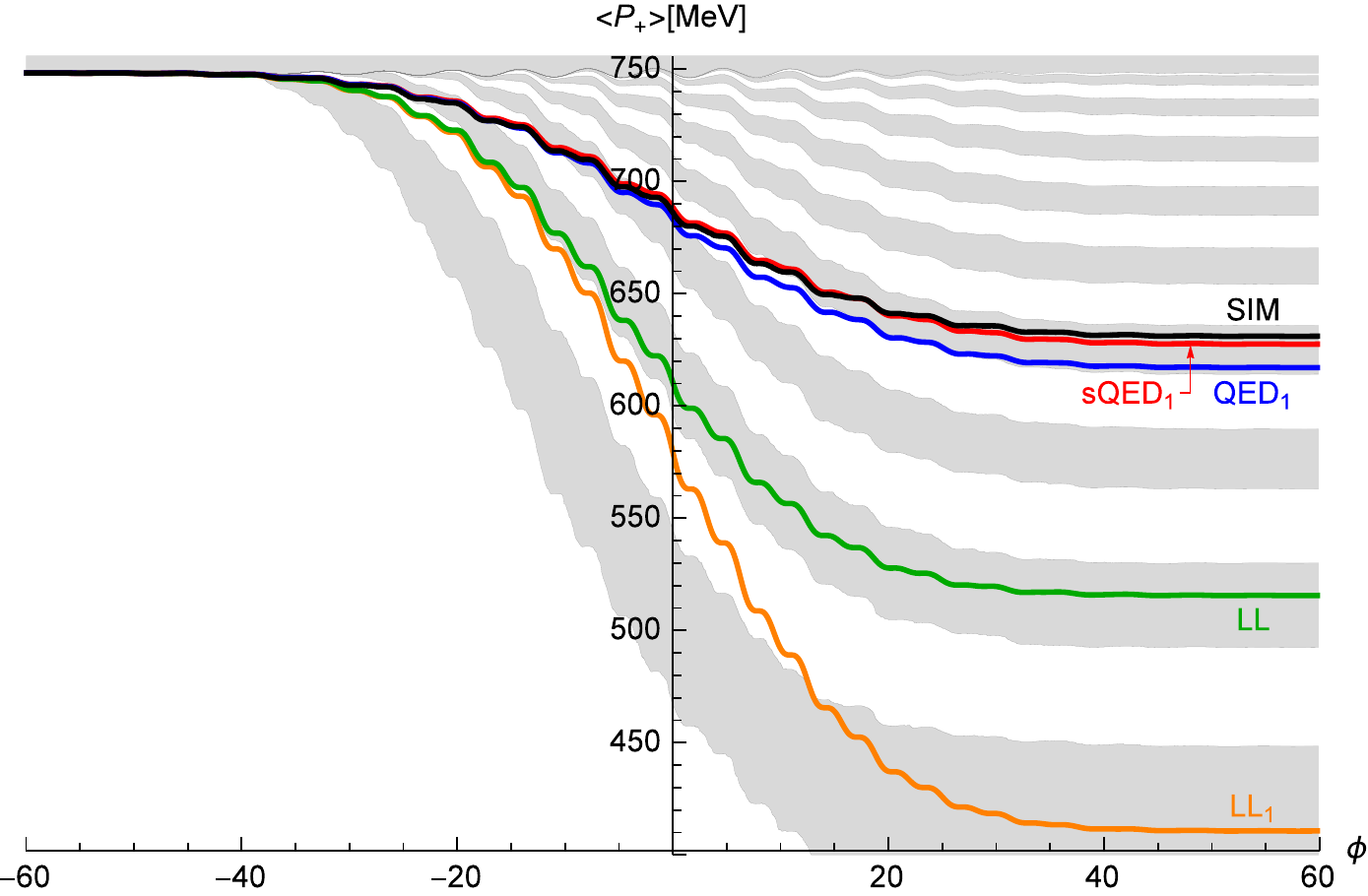}
	\includegraphics[width=\columnwidth]{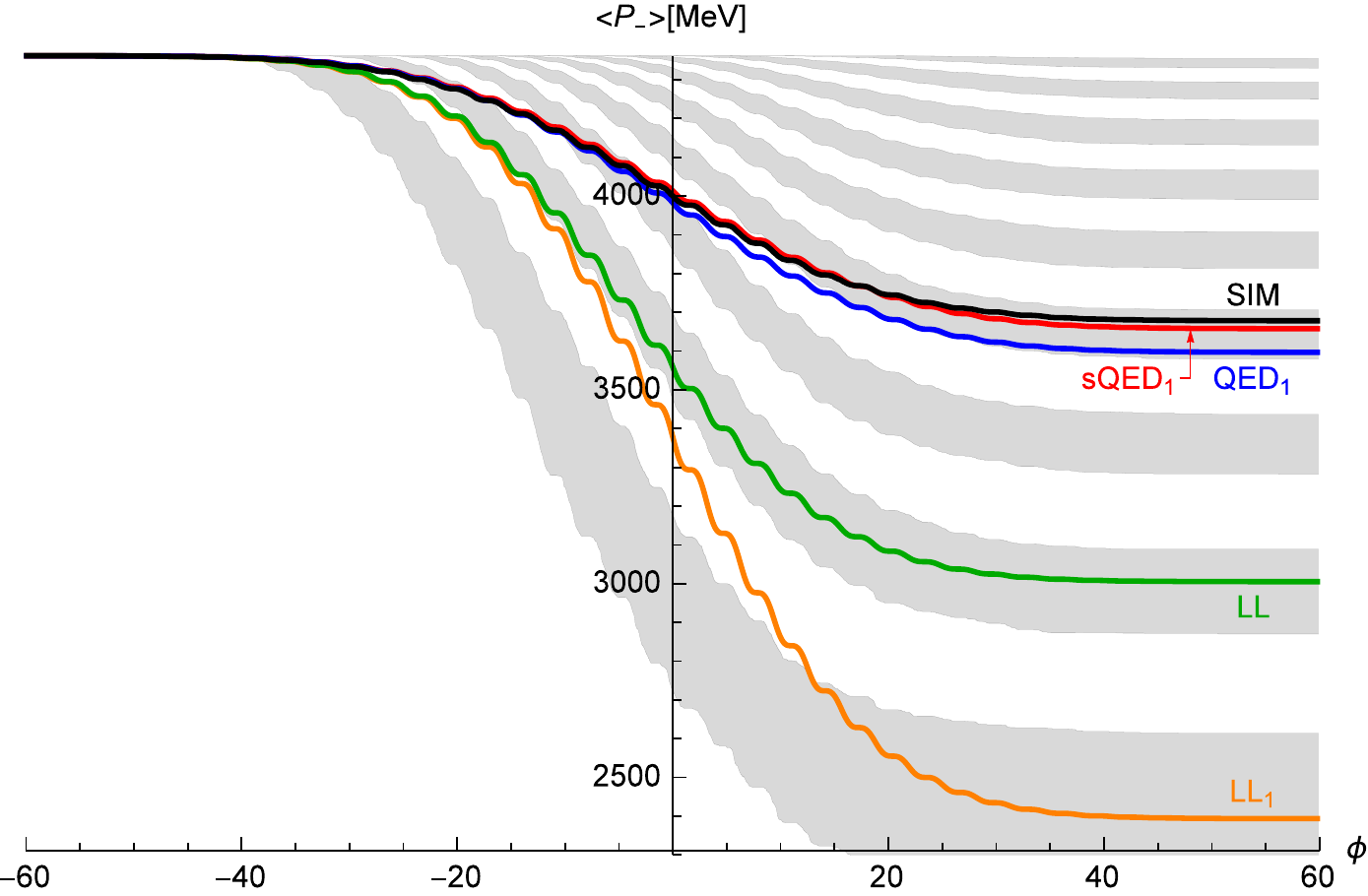}
	\includegraphics[width=\columnwidth]{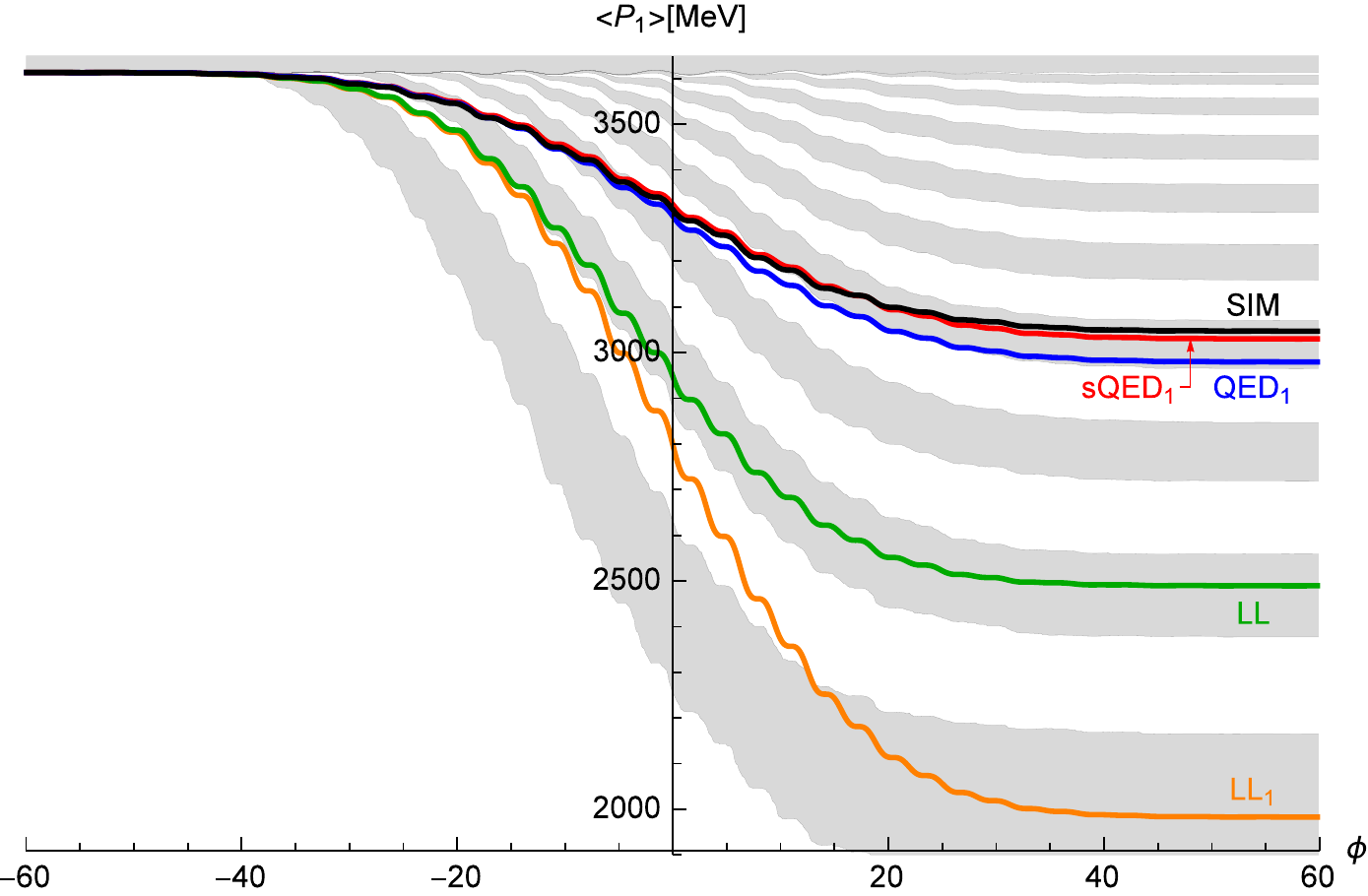}
	\caption{\label{FIG:GEMINI45} All nontrivial electron momentum components for a $45^\circ$ collision in the crossover regime, colours as in the Letter.  $a_0=10$ and $\gamma=10^4$. A little over $60\%$ of trajectories show lower-than-average energy loss.}
\end{figure}
\newpage

\onecolumngrid

\end{document}